# Effect of particle size on the phase transformation behavior and equation of state of Si under hydrostatic loading

Sorb Yesudhas[1]*, Valery I. Levitas[1,2,3]*, Feng Lin[1], K. K. Pandey[4], and Maddury Somayazulu[5]

[1]Department of Aerospace Engineering, Iowa State University, Ames, Iowa 50011, USA

[2]Department of Mechanical Engineering, Iowa State University, Ames, Iowa 50011, USA

[3]Ames National Laboratory, Iowa State University, Ames, Iowa 50011, USA

[4]High Pressure & Synchrotron Radiation Physics Division, Bhabha Atomic Research Centre, Mumbai 400085, India

[5]HPCAT, X-ray Science Division, Argonne National Laboratory, Argonne, Illinois 60439, USA

*Corresponding authors. Email: sorbya@iastate.edu, vlevitas@iastate.edu

High-pressure synchrotron X-ray diffraction (XRD) studies have been conducted on three types of Si particles (micron, 100 nm, and 30 nm). The pressure for initiation of Si-I→Si-II phase transformation (PT) essentially increases with a reduction in particle size. For 30 nm Si particles, Si-I directly transforms to Si-XI by skipping the intermediate Si-II phase, which appears during the pressure release. The evolution of phase fractions of Si particles under hydrostatic compression is studied. The equation of state (EOS) of Si-I, Si-II, Si-V, and Si-XI for all three particle sizes is determined, and the results are compared with other studies. A simple iterative procedure is suggested to extract the EOS of Si-XI and Si-II from the data for a mixture of two and three phases with different pressures in each phase. Using previous atomistic simulations, EOS for Si-II is extended to ambient pressure, which is important for plastic strain-induced phase transformations. Surprisingly, the EOS of micron and 30 nm Si are identical, but different from 100 nm particles. In particular, the Si-I phase of 100 nm Si is less compressible than that of micron and 30 nm Si. The reverse Si-V→Si-I PT is observed for the first time after complete pressure release to the ambient for 100 nm particles.

## I. Introduction

Silicon is the second most abundant element on the Earth's crust after oxygen. The outstanding electronic properties of this covalent semiconductor ($E_g$ = 1.1 eV) find applications in information technology, photovoltaics, nano-electromechanical systems (MEMS/NEMS), etc[1-5]. Si has been considered a promising platform with tremendous potential for advancing scaled-up quantum computing[5]. Recent studies reveal that Si-III is a direct ultra-narrow band gap semiconductor with a band gap of 30 meV, which can be used in infrared plasmonic devices[6]. Silicon undergoes multiple polymorphic phase transformations under high pressure[7-12]. The scaling down of Si from a micron to nm size shows extraordinary electronic, optical, and mechanical properties[13-16].

With hydrostatic compression, the ambient cubic diamond Si-I phase (S.G: $Fd\bar{3}m$, Z=8) undergoes a pressure-driven cubic to tetragonal Si-II (S.G: $I4_1/amd$, Z = 4) reconstructive type phase transformation (PT) at around 11 GPa as a result of stretching of Si-1 unit cell along [100], [010] and compression along [001][7-12] with a significant volume collapse of 23.7%. This PT is accompanied by a semiconductor-to-metal transition[7-12]. With further compression, the antiparallel displacements of atoms in the Si-II lattice along the crystallographic $c$-axis by ±0.068 induce the distortion of the tetragonal Si lattice to another lower symmetry orthorhombic Si-XI (S.G: Imma, Z=4) lattice[7,11]. This transition involves no coordination number change but is coupled with the Wyckoff position change from 4($a$) to 4($e$)[7,11]. Similarly, the antiparallel displacements of Si atoms in the orthorhombic lattice along the $c$-axis by ± 0.057$c$ bring about orthorhombic to another displacive type simple hexagonal phase Si-V (S.G: P$6$/mmm, Z=1)[7]. The coordination number of Si-I, Si-II, Si-XI, and Si-V phases are 4, 6, 6, and 8, respectively. Si exhibits quite different structural behavior during unloading. With fast unloading from Si-II, two tetragonal phases are obtained: Si-VIII (S.G: P$4_12_12$) and Si-IX (S.G: P$422$)[17]. When Si-II is unloaded slowly using a liquid pressure transmitting medium, a rhombohedral semiconducting phase Si-XII (S.G: R$\bar{3}$), is observed at 9.9 GPa, which coexists with the Si-III (S.G: Ia$\bar{3}$) (Si-XII+Si-III) below 3.2 GPa, and is in turn almost transform to Si-III at 2 GPa[18-19]. A small portion of Si-XII was noticed even after the complete pressure release[18-19].

When a material is compressed in a DAC or compressed and sheared in a rotational DAC without a pressure-transmitting medium, the material undergoes plastic strain-induced PTs, which show different behavior in comparison with the PTs observed under hydrostatic conditions[20-23]. The initiation of phase transformation pressure for samples subjected to quasi-hydrostatic compression is determined by the preexisting defects, like dislocations and grain boundaries. In contrast, the creation of new defects during the plastic flow with much stronger stress concentrators, like dislocation pileups with a large number of dislocations, is responsible for a significant reduction in PT pressure for plastic strain-induced phase transformation[21-23]. The determination of pressure in plastic strain-induced PTs study requires an EOS for each phase of the material, which is determined under hydrostatic conditions. The actual (average) pressure at a radial point ($r$) on the sample will be the sum of the product of the pressure in each phase multiplied by corresponding volume fractions. The radial pressure distribution in a sample compressed in a DAC or compressed and sheared in a rotational DAC is very heterogeneous[20,24-27]. It is determined using the measured radial distribution of the volume of phases and applying EOS of these phases determined under hydrostatic conditions. In particular, silicon exhibits quite complex PT sequence along the sample diameter under non-hydrostatic compression and torsion, and calculating pressure in the Si phases at multiple points along the sample diameter for various pressures requires accurate EOS of different Si phases with $P$-$V$ data from the pure Si phases[20]. With plastic compression and shear, we reported a drastic pressure reduction of Si-I→Si-II PT from 13.5 GPa to 2.6 GPa for micron Si, and it is from 16.2 GPa to 0.3 GPa for 100 nm Si[20]. Thus, EOS for Si II (and other phases) should be robustly extrapolated for pressures outside the region of the existence of phases under hydrostatic loading. The only way to do this is to utilize theoretical (first principle) results, which we will utilize in the current paper.

Despite several high-pressure PT studies reported for various grain sizes of silicon[10, 13-15, 28-29], a detailed comparative high-pressure structural behavior of different grain sizes of Si is missing. Using Le Bail fitting of synchrotron XRD data, the equation of state of micron size Si up to one megabar pressure has been determined recently[28]. In addition, the phase transition behavior of silicon with different morphology and size has been investigated[29]. However, this study did not provide a detailed XRD analysis to extract the lattice parameters, volume of high-pressure phases, and phase fractions, which are crucial to understanding the quantitative evolution of high-pressure phases[30]. Also, we are not aware of any high-pressure studies that reported phase fractions of high-pressure phases of Si, and most of the studies utilized Le Bail fitting on XRD data, which led to inaccuracy in the determination of lattice parameters and, hence, the volume of high-pressure phases. The preferred orientation effects on high-pressure phases of Si pose difficulties in extracting phase fractions. Also, the experimental EOS of Si-II is not available in the literature because of the narrow pressure range of its existence, and it always coexists with other phases. Without knowing the volume fraction of each phase, it is impossible to extract the pressure and EOS of each phase in the mixture.

In this work, we performed a first detailed comparative study of the structural behavior of silicon with three different particle sizes, 1 μm, 100 nm, and 30 nm, under hydrostatic compression with high-pressure synchrotron X-ray diffraction technique. The volume and phase fraction of Si phases are extracted using the Rietveld refinement method. A known theoretical EOS for Si-II is used. In such a way, the EOS of different Si phases with different particle sizes were studied and compared with other reports, if available.

## II. Experimental Details

Single crystal, 100 nm, and 30 nm Si particles were used for the high-pressure studies. The single crystal silicon sample was purchased from Sigma-Aldrich (CAS NO.: 7440-21-3) and was powdered to a ~1 μm fine powder by grounding with a pestle and mortar. The Si nano samples with a particle size of < 100 nm (TEM) and purity of ≥ 98 % were purchased from Sigma-Aldrich (CAS No.: 7440-21-3). The Si nanopowders with a particle size of 30 nm were purchased from Meliorum Technologies, Inc. High-pressure experiments were carried out using symmetric type diamond anvil cells (DACs) and SSDAC (from DAC Tools, LLC) with culet diameters of 300 and 400 μm. The micron and 100 nm samples were compacted by pressing the samples between two stainless steel (S. S.) gaskets. A small pellet of the compacted sample was placed into the S. S. sample chamber. The 30 nm Si nanopowder was slightly pressed using a hydraulic press to make a thin pellet. The He PTM was used to achieve hydrostatic conditions[31]. The pressure in the PTM and sample was determined by the ruby fluorescence method[32]. All the high-pressure synchrotron XRD experiments were carried out at the 16-ID-B beam line, HPCAT, utilizing Advanced Photon Source with X-ray wavelengths, 0.4246 Å and 0.3445 Å and the X-ray beam with a spot size of 5μm x 4μm. The 2D XRD image was converted to a 1D pattern using dioptase software[33]. The XRD patterns were refined using GSAS-II software to extract the lattice parameters and volume fractions of Si phases using the Rietveld refinement method[34]. The spherical harmonics texture model was used to fit the high-pressure Si phases. The bulk modulus and its first derivative were calculated using the EOSFIT-7 GUI program[35].

## III. Results and Discussions

## III.1. Phase transformation behavior

High-pressure XRD patterns of micron size Si for representative pressures and the pressure dependence of the volume/Z are shown in Figs. 1a, b. The Si-I phase is stable up to ~13.3 GPa, and at ~13.5 GPa, it undergoes a pressure-driven symmetry lowering structural phase transformation into a mixture of tetragonal Si-II and an orthorhombic Si-XI (Imma) phases (Figs. 1a, b). The XRD pattern at 13.5 GPa can be fitted with all these three phases (Si-I, Si-II, and Si-XI), suggesting the observed transition is a mixed-phase PT, and these phases coexist up to 14.1

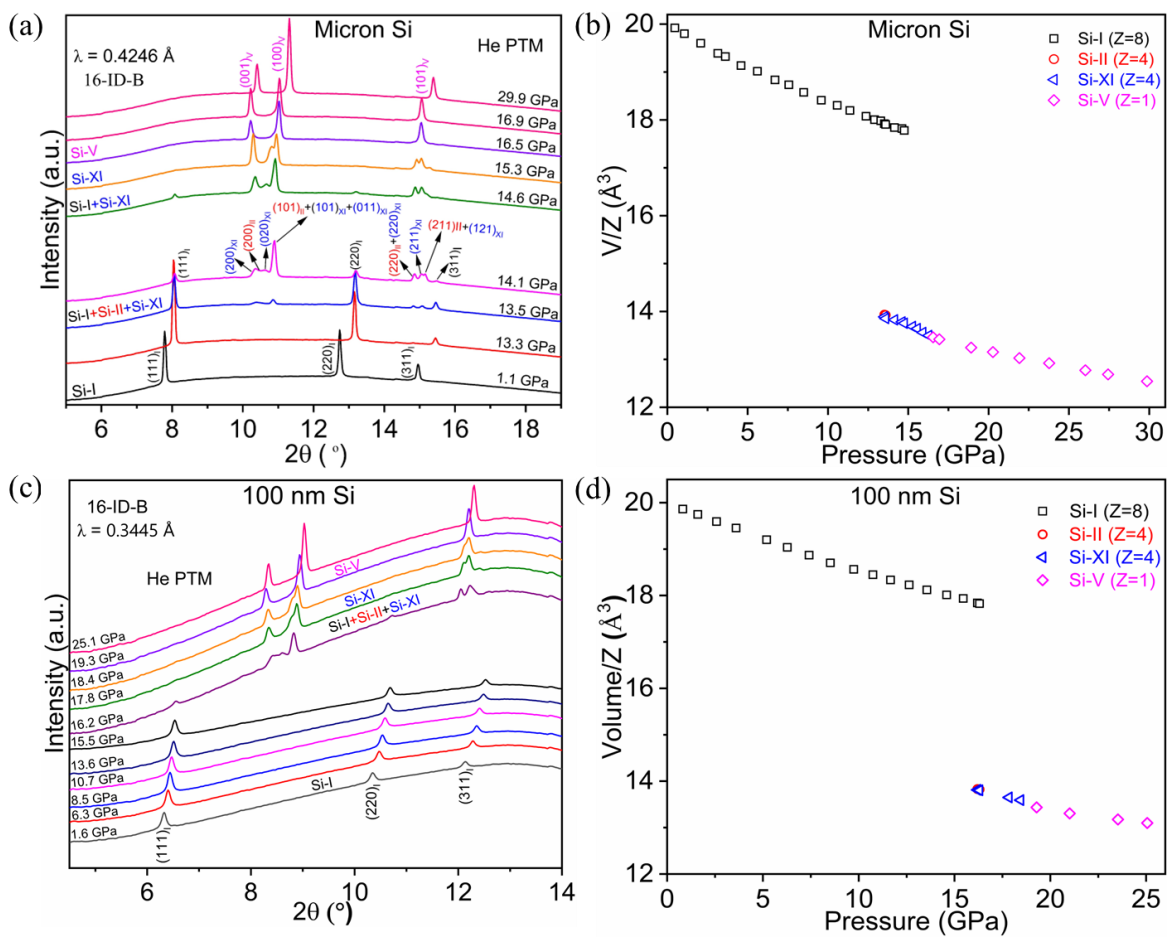

**Fig. 1. Pressure-induced phase transformations in micron and 100 nm Si. (a)** High-pressure XRD patterns of micron size Si at various pressures. The Si-I, Si-II, and Si-XI mixed phases are indexed at 14.1 GPa. The subscripts I, II, XI, and V of the (hkl) Bragg planes represent the Bragg planes of the Si-I, Si-II, Si-XI, and Si-V phases, respectively. **(b)** Volume/Z *vs.* pressure for micron size Si. The error bar is excluded from all the figures if it overlaps with the symbols. **(c)** High-pressure XRD patterns of 100 nm Si for selected pressures. **(d)** Volume/Z *vs.* pressure of 100 nm Si.

GPa (Figs. 1a, 1b). The enhancement of the intensity of the XRD peaks of the Si-II and Si-XI lattices and a dramatic drop in that of the Si-I phase evidence PT of the Si-I lattice into Si-II and Si-XI lattices. At 14.6 GPa, a portion of Si-I and the remnant of the Si-II lattices transform into the Si-XI lattice, and the Si-I is seen to coexist with Si-XI (Figs. 1a, b). With further pressure increase, the remaining Si-I lattice transforms into Si-XI lattice at 15.3 GPa (Figs. 1a, b). At 16.5 GPa, a pure Si-V phase is formed, which is stable up to the maximum pressure achieved, 29.9 GPa. The Rietveld refinement of the Si-V phase for the micron Si at 23.8 GPa is shown in Fig. A1. The Si-I→Si-II PT is associated with a volume collapse of 22.3 %, whereas the Si-II→Si-XI and Si-XI→Si-V PTs are continuous (Figs.1a, b). Our results on micron Si agree with those obtained in[28]. The observed onset of Si-I→Si-I+Si-II+Si-XI PT at 13.5 GPa agrees with the initiation of PT in[28]. When the pressure in the sample is completely released to the ambient, a coexistence of Si-III and Si-XII phases is discerned.

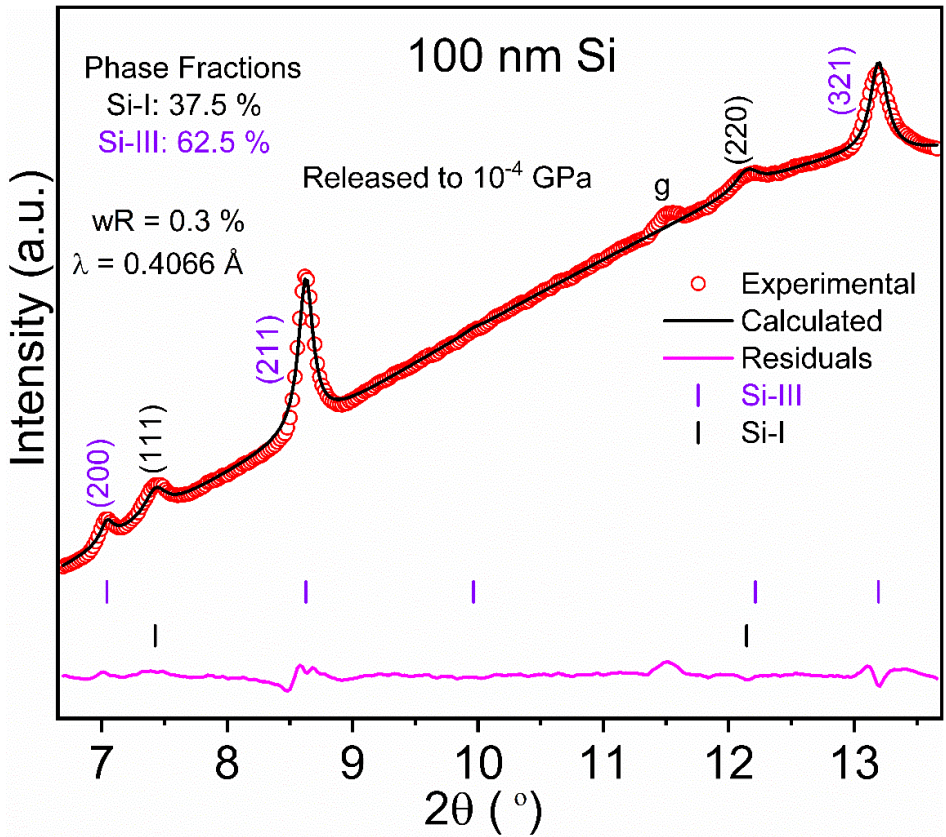

**Fig. 2. The first observation of reverse Si-V→Si-I PT in hydrostatic high-pressure experiments.** Rietveld refinement of 100 nm Si after releasing to the ambient from Si-V at 25.1 GPa. Si-I phase is coexisting with Si-III. 'g' is the gasket peak.

Fig. 1c shows the high-pressure XRD pattern of 100 nm Si for selected pressures. The PT sequence in 100 nm Si is similar to that in micron Si, except for the initiation of PT pressures of high-pressure phases. As the particle size is reduced from micron to 100 nm Si, the Si-I transformed into a mixture of Si-I, Si-II, and Si-XI at 16.2 GPa, which is 2.7 GPa higher as compared to the initiation of PT pressure obtained in micron Si (Figs. 1c, d and Fig. A2). The Si-I, Si-II, and Si-XI phases are seen to coexist in a very narrow pressure range. At 17.8 GPa, the Si-I and Si-II lattices completely transform into Si-XI (Figs. 1c, d). The Si-XI→Si-V PT completes at 19.3 GPa, stable up to the maximum pressure of 25.1 GPa achieved in the experiment (Figs. 1c, d). When the pressure is completely released to the ambient, Si-I (37.5 %) and Si-III (62.5 %) phase mixtures are obtained (Fig. 2). The reverse Si-V→Si-I PT is observed for the first time in the hydrostatic compression-decompression experiments (Fig. 2).

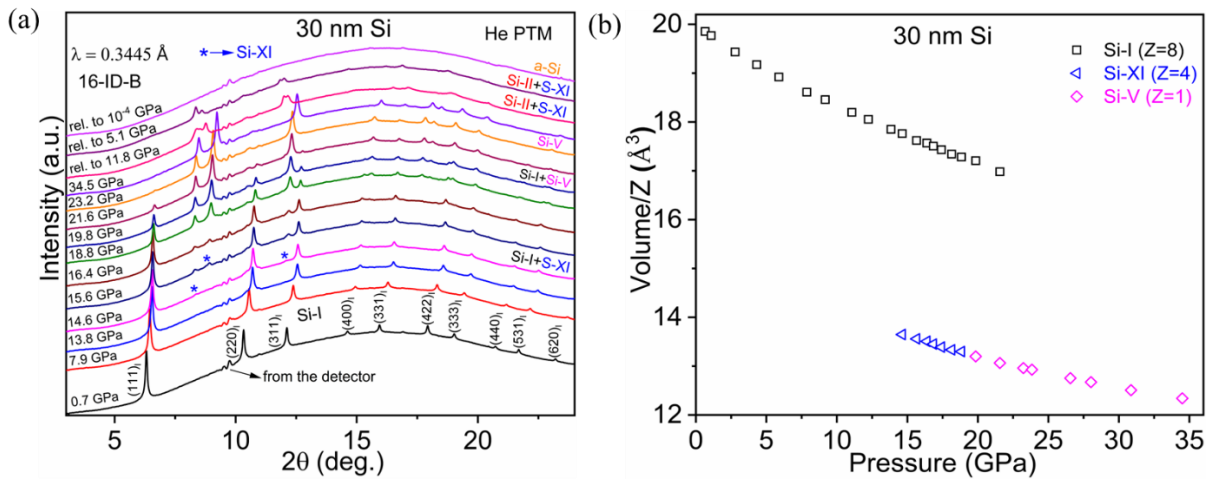

**Fig. 3. Pressure-induced phase transformation in 30 nm Si.** **(a)** High-pressure XRD patterns of 30 nm Si at various pressures. **(b)** Volume/Z *vs.* pressure of 30 nm Si.

In contrast to micron and 100 nm Si particles, the 30 nm Si particles start directly transforming into a Si-XI phase at 14.6 GPa by skipping the intermediate Si-II phase (Figs. 3a, b). The XRD patterns up to 18.8 GPa can be fitted well with Si-I and Si-XI phase coexistence (Si-I+Si-XI). The XRD pattern at 18.8 can be fitted to both Si-I+Si-XI and Si-I+Si-V phase coexistences, but we obtained the best fit with Si-I+Si-XI phase mixture. At 19.8 GPa, the Si-V phase appears, which coexists with Si-I. The Si-I phase coexists with Si-V up to 23.2 GPa. Note that perfect Si-I crystal shows transverse acoustic phonon instability at 18.3 GPa, according to the first principle simulations[36], i.e., it cannot exist above this pressure. Thus, Si-I is stabilized by size and surface effects. After pressure release to ambient, amorphous Si is observed, like in[29,38]. The Si-I→Si-XI PT is accompanied by a volume collapse of 23.2 %, whereas the Si-XI→Si-V PT is practically continuous. For micron and 100 nm Si, the Si-II and Si-XI phases coexist over a very narrow pressure range of < 1 GPa. For 30 nm Si, the Si-I coexists with Si-XI (Si-I+Si-XI) and Si-V (Si-I+Si-V) over an extensive pressure range of about 7 GPa (Fig. 3).

The initiation pressure for the Si-I→Si-II/Si-XI PT increases from 13.5 GPa to 16.2 GPa for 100 nm Si and is 14.6 GPa for 30 nm Si (Figs. 1c, d, and 2). A pure Si-XI phase is observed for micron and 100 nm Si at 15.3 GPa and 18.1 GPa, respectively, whereas the Si-I phase always coexists with the Si-XI phase for 30 nm Si (Figs. 1 and 2). The initiation pressure for Si-XI→Si-V increases from micron to 30 nm Si (16.5 GPa for micron Si, 19.3 GPa for 100 nm Si, and 19.8 GPa for 30 nm Si). The elevation of Si-I→Si-II PT pressure upon reducing the particle size is due to the reduction in the number of stress concentrators (dislocations and grain boundary effects) and contribution from the surface energy. The surface energy strongly increases during Si-I→Si-II PT, suppressing this PT for all particles. This effect increases with a reduction in particle size, and for 30 nm Si, Si-XI and V appear before Si-II due to their smaller surface energy.

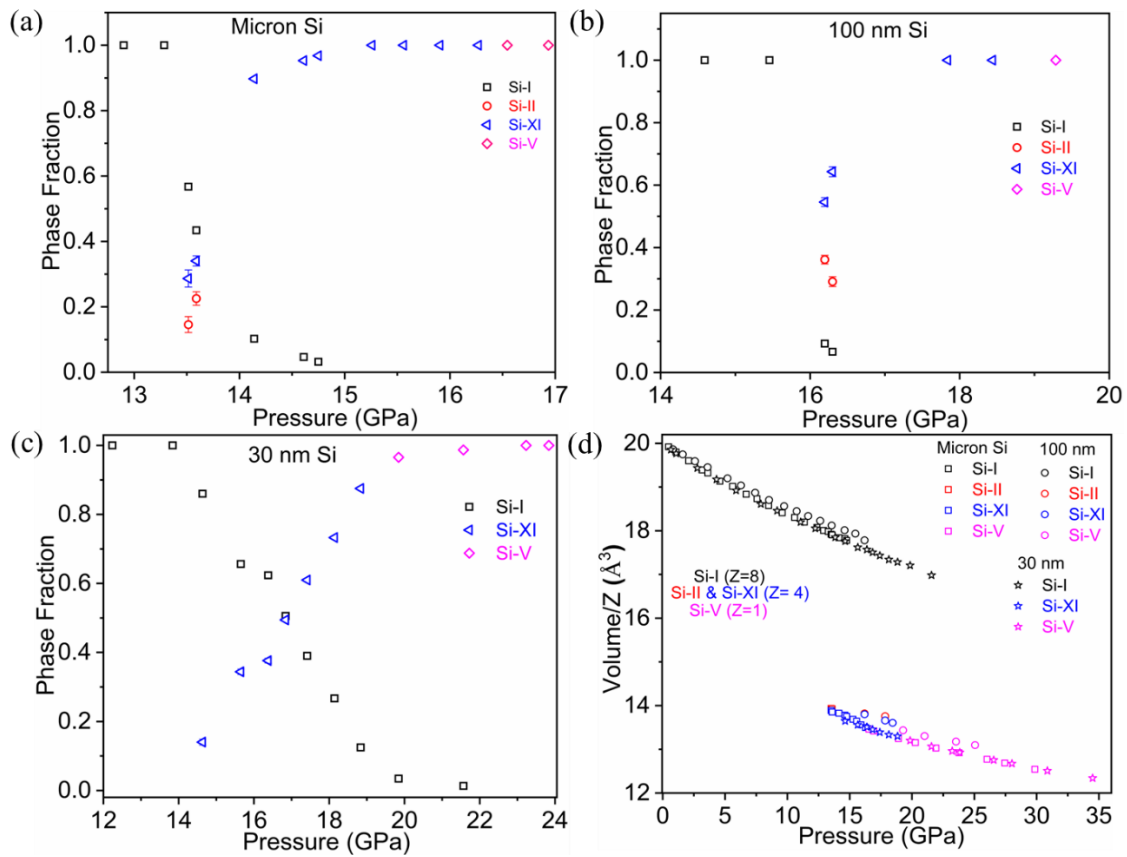

**Fig. 4. Evolution of phase fraction and volume of Si phases under pressure. (a)** Phase fraction *vs.* pressure of micron Si. **(b)** Phase fraction *vs.* pressure of 100 nm Si. **(c)** Phase fraction *vs.* pressure of 30 nm Si, and **(d)** a comparison of volume/Z of Si particles *vs.* pressure.

These results raise theoretical challenges for their quantitative descriptions. Xuan *et al.*[29] claimed that the compression of Si nanoparticles with and without a pressure-transmitting medium will not alter the phase-transition sequence. However, in our high-pressure studies on 30 nm Si particles, the non-hydrostatic compression alters the phase transformation sequence and drastically reduces the onset of Si-I→Si-II PT pressure[20]. With He PTM, Si-I directly transforms to Si-XI by skipping Si-II, whereas we observed Si-II at 4.9 GPa for 30 nm Si under non-hydrostatic compression[20].

The phase fraction and volume versus pressure for all three types of Si particles are shown in Figs. 4a-d. For micron Si, the maximum phase fraction of Si-II observed is 23 % at 13.6 GPa, whereas it is 36 % at 16.2 GPa for 100 nm Si. The coexistence range of the Si-II phase with Si-I and Si- XI is very narrow for micron and 100 nm Si particles, and Si-II does not appear for 30 nm Si. The V/Z of Si-I, Si-XI, and Si-V phases of micron and 30 nm Si exhibit an identical behavior (Fig. 4d). The phase fractions of Si-I and Si-XI phases in 30 nm Si particles at 14.6 GPa are 14 % and 86 %, respectively. Similarly, the phase fractions of Si-I and S-V at 19.8 GPa are 3.4 % and 96.6 %, respectively. During pressure release, Si-V transforms to a phase mixture of Si-II and Si-XI, which is stable down to 5.1 GPa. When the pressure is completely released to the ambient, it transforms to the amorphous phase, which agrees with[29,38].

### III.2. Equation of state of Si phases

#### A. EOS for Si-I and Si-V

The volume at P = 0 ($V_0$), bulk modulus ($B_0$), and its first derivative ($B_0'$) of Si phases of micron, 100 nm, and 30 nm Si are calculated by fitting *P-V* data of corresponding phases to the Murnaghan, Birch-Murnaghan, and Vinet EOS[39-40] (Figs. 5, 6), and the results are shown in Table 1. The bulk moduli values of the present work are compared with the most recent study in[28] (Table 1). Our $B_0$ and $B_0'$ values of Si-I for micron Si agree with the values reported in other references[9,28,41] (Table 1). Moreover, the $B_0$ and $B_0'$ values of the Si-I phase agree with the values obtained by the ultrasonic measurements ($B_0$ = 97.88 GPa and $B_0'$ = 4.23)[41]. The volume at zero pressure $V_0$ of the Si-V phase is taken from the theoretical prediction[42] (15.204 Å$^3$). Figs. 5 and 6 demonstrate a good correspondence between analytical EOS and experimental data for EOS of Si-I and Si-V for all particles. The bulk moduli values of micron Si-V are slightly underestimated in[28] (Table 1), which is evident from the steepness of the *P-V* curve in Fig. 5b. Our values are more accurate because we used pure *P-V* data for Si-V, and we also calculated the volume using Rietveld refinement method. The Si-I phase of 100 nm Si is found to be less compressible when compared with micron and 30 nm Si (Fig. 4d, Table 1). The compressibility of ambient and high-pressure phases of 30 nm Si is comparable to the micron Si (Fig. 4d, Tables 1 and 2). In Fig 3d, the behavior of V/Z *vs.* P of micron Si is akin to the 30 nm Si; hence, all the equations of states of micron Si can be used for 30 nm Si as well. Such a non-monotonous dependence of the EOS for Si-I on the particle size is unexpected and requires theoretical analysis. The EOS of Si-III was obtained in[37] (Table 2).

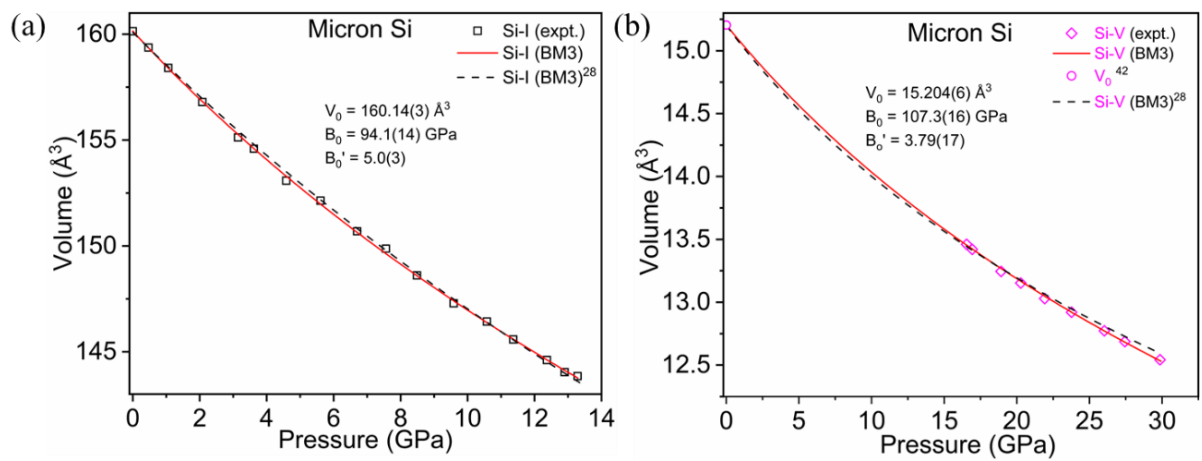

**Fig. 5.** Calculating bulk modulus and its derivative of **(a)** Si-I phase of micron Si. The dotted line represents the simulated EOS taken from[28]. Volume *vs.* P of Si-I phase of micron Si. **(b)** Volume *vs.* P of Si-V phase of micron Si. V₀ is taken from[42]. The dotted line is the simulated pattern from[28].

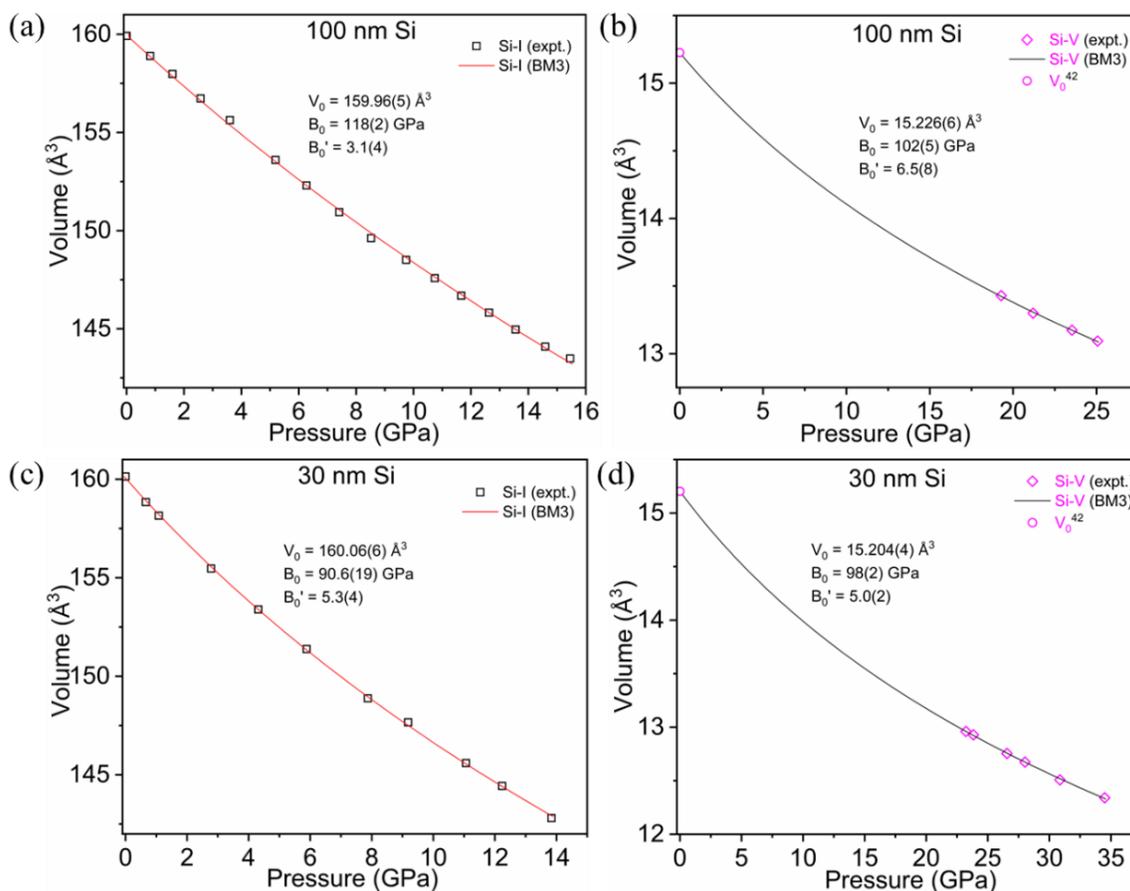

**Fig. 6.** EOS of **(a)** Si-I phase of 100 nm Si **(b)** Si-V phase of 100 nm Si. $V_0$ is taken from theory[42]. **(c)** Si-I phase of 30 nm Si **(d)** Si-V phase of 30 nm Si. $V_0$ is taken from theory[42].

| EOS | | Micron Si | | 100 nm Si | | 30 nm Si | |
|---|---|---|---|---|---|---|---|
| | | Si-I | Si-V | Si-I | Si-V | Si-I | Si-V |
| 3$^{rd}$ order Birch-Murnaghan | $V_0$ (Å$^3$) $B_0$ (GPa) $B'_0$ | 160.14(3) 94.1(14) 5.0(3) | 15.204(6) 107.3(16) 3.79(17) | 159.96(5) 118(2) 3.1(4) | 15.226(6) 102(5) 6.5(8) | 160.06(6) 90.6(19) 5.3(4) | 15.204(4) 98(2) 5.0(2) |
| 3$^{rd}$ order Birch-Murnaghan[28] | $V_0$ (Å$^3$) $B_0$ (GPa) $B'_0$ | 160.088(8) 101.5(3) 3.45(5) | 15.2(1) 99(10) 5.0(6) | …… | …….. | …… | …… |
| 2$^{nd}$ order Birch-Murnaghan | $V_0$ (Å$^3$) $B_0$ (GPa) $B'_0$ | 160.11(3) 98.5(5) 4.0 | 15.206(6) 105.5(5) 4.0 | 160.00(5) 113.5(8) 4.0 | 15.225(13) 120.7(12) 4.0 | 159.94(7) 96.9(8) 4.0 | 15.203(9) 107.1(7) 4.0 |
| Murnaghan | $V_0$ (Å$^3$) $B_0$ (GPa) $B'_0$ | 160.14(3) 94.5(14) 4.7(3) | 15.204(6) 108.6(16) 3.47(16) | 159.96(5) 119(2) 2.9(4) | 15.226(5) 105(4) 5.6(5) | 160.05(6) 91.3(19) 4.9(3) | 15.204(4) 100.0(18) 4.44(17) |
| Vinet | $V_0$ (Å$^3$) $B_0$ (GPa) $B'_0$ | 160.14(3) 93.9(14) 5.1(3) | 15.204(6) 106.8(18) 3.9(2) | 159.96(5) 119(2) 3.1(4) | 15.226(6) 102(5) 6.5(7) | 160.06(6) 90.4(19) 5.4(4) | 15.204(4) 96(2) 5.3(2) |

**Table 1.** Volume at zero pressure $V_0$ (Å$^3$), the bulk modulus $B_0$ (GPa) and its first derivative ($B'_0$) of Si-I and Si-V phases of micron, 100 nm and 30 nm Si are calculated by fitting our experimental *P-V* data to the Murnaghan, second and third order Birch-Murnaghan, and Vinet EOS. The parameters of the third-order Birch-Murnaghan EOS from[28] are compared with the values obtained by our data.

**B. Extracting EOS for Si-II and Si-XI phases from phase mixtures**

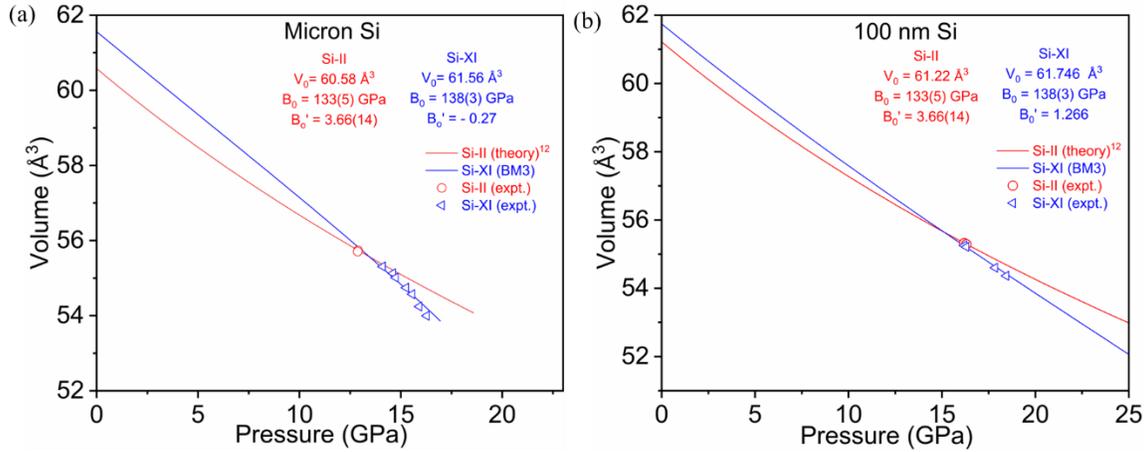

**Fig. 7.** EOS of Si-II and Si-XI phases of **(a)** pure micron Si and **(b)** 100 nm Si. The experimental data points for Si-II and Si-XI are shown in circles and triangles, respectively.

For micron Si XI, two experimental *P-V* data points of Si-XI coexist with Si-I and Si-II (Fig. 1); three data points coexist with Si-I; and there are four points for a single-phase Si-XI. For the coexisting of Si-I and Si-XI phases, as we know the pressure $p_I$ and phase fraction $c_I$ of pure micron Si-I (using pure Si-I EOS), the phase fraction $c_{XI}$ of Si-XI, and the pressure $P$ in the phase mixture measured by ruby. The pressure in pure Si-XI is obtained using the phase mixture rule,

$$P = c_I P_I + c_{XI} P_{XI} . \qquad (1)$$

This gives us *P-V* data for three more points, total 7 points for pure Si-XI, in a broader pressure range, to fit BM3 EOS for Si-XI (Table 2 and Fig. 7). It is to be noted that, even under non-hydrostatic compression, Si-XI was not observed below 10 GPa for any particles, extrapolation below this value does not have practical importance[20]. The same procedure was done for 100 nm Si particles. EOS for 30 nm Si-XI is the same as for micron particles.

Since we acquired only two data points for Si-II coexisting with Si-I and Si-XI for micron and 100 nm Si, we used the theoretical *P-V* data from[12] for the Si-II EOS. First, using the phase mixture rule for a three-phase mixture

$$P = c_I P_I + c_{II} P_{II} + c_{XI} P_{XI} \qquad (2)$$

with all known parameters but the pressure in Si-II $P_{II}$, we calculate $P_{II}$ for two experimental points. $P_{XI}$ is obtained by using Si-XI EOS (Fig. 8 and Table 2). According to[12], the Si-II is metastable at normal pressure, i.e., it can exist if retained instead of transforming to Si-XII and Si-III, which was confirmed experimentally for some sample points in[20]. That is why interpolation down to zero pressure for Si-II is crucial.

The results are presented in Table 2. First, we determined EOS for Si-XI for micron and 100 nm Si based on our experimental data. Then, the theoretical data from[12] is described by the third order Birch Murnaghan EOS with Vo = 60.29 (14) Å$^3$, $B_0$ =133(5) GPa and $B_0'$ =3.66(14) GPa.

| EOS | | Micron Si | | | 100 nm Si | |
|---|---|---|---|---|---|---|
| | | Si-II | Si-XI | Si-III | Si-II | Si-XI |
| 3rd order Birch-Murnaghan (This study) | $V_0$ (Å$^3$) $B_0$ (GPa) $B_0'$ | 60.58 133(5) 3.66(14) | 61.56 138(3) -0.27 | ……. | 61.22 133(5) 3.66(14) | 61.746 138(3) 1.266 |
| 3rd order Birch-Murnaghan (Si-II[12], Si-XI[28,42] and our data) | $V_0$ (Å$^3$) $B_0$ (GPa) $B_0'$ | 60.29(14) 133(5) 3.66(14) | 62.89(3) 98(3) 1.9(4) | ……. | ……. | …… |
| 2nd order Birch-Murnaghan (Si-III[37]) | $V_0$ (Å$^3$) $B_0$ (GPa) $B_0'$ | ……. | 68.52(24) 42(6) 4(1) | 292.36 117.0 4.0 | ……. | …… |

**Table 2**. The BM3 EOS of Si-XI and Si-II phases of micron and 100 nm Si from our experimental *P-V* data with a mixture of two and three phases are obtained by using simple iterative procedure. The EOS for Si-II is also approximated by fitting simulated *P-V* data in[12] to the BM3 EOS. The EOS of Si-XI is also obtained by combining our data with data from[28] and one of the theoretical values for Vo from[42]. EoS for Si-III is taken from[37]

For micron particles, we combined our experimental points with theoretical data from[12], keeping the same $B_0 =133(5)$ GPa and $B'_0 =3.66(14)$ GPa, and obtained slightly higher $V_0 = 60.58$ Å$^3$. A slightly higher Vo than in[12] is reasonable because calculations in[12] are at 0 K and there is a thermal expansion while heating to room temperature.

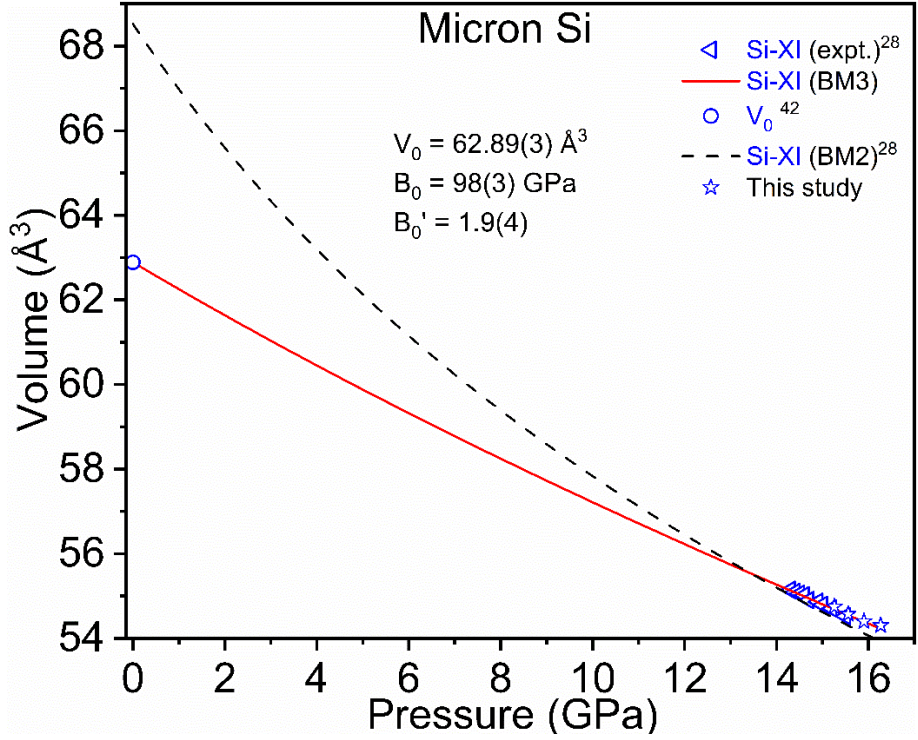

**Fig. 8.** Si-XI EOS of micron Si. The combined *P-V* data of pure Si-XI phase of micron Si from this work and also from[28] is fitted to the BM3 EOS with $V_0$ = 62.89 Å$^3$, which is one of the theoretical values from[42]. Triangle and star symbols represent the pure Si-XI *P-V* data from[28] and from our work.

For 100 nm Si, two experimental points have slightly higher pressure (3.3 and 3.4 GPa) than for micron particles. That is why we keep the same $B_0$ and $B'_0$, and slightly increase Vo to 61.22(14) Å$^3$. Then, we slightly refine Vo and $B'_0$ for Si-XI to get approximately continuous *P-V* curve for Si-II to Si-XI transition at transition pressure (Fig. 7 and Table 2).

We have also calculated the EOS for micron Si-XI by combining our pure Si-XI *P-V* data with experiments from[28]. We also utilized one of the theoretical values of $V_0$ = 62.89(3) Å$^3$ from[42]. The obtained EOS is compared in Fig. 8 with the BM2 EOS reported in[28] where a large value of $V_0 = 68.52$ Å$^3$ was chosen[12]. Moreover, in[28], the authors used both mixed and pure Si-XI P-V data to obtain BM2 EOS which is why the EOS from[28] shows a deviation from our Si-XI P-V data.

## IV. Conclusions

A comparative study of high-pressure PT behavior of three types of Si viz, micron, 100 nm, and 30 nm Si is performed. The pressure for initiation of Si-I→Si-II phase transformation (PT) essentially increases with a reduction in particle size. The Si-II and Si-XI phases of micron and 100 nm Si appear from Si-I simultaneously and coexist over a narrow pressure range of < 1 GPa, whereas for 30 nm Si, Si-I directly transforms to Si-XI (skipping Si-II), and Si-I coexists with Si-XI (Si-I+Si-XI) and Si-V (Si-I+Si-V) over a wide range of pressures. The increase in Si-I→Si-II PT pressure with reduction in the particle size can be rationalized by the reduction in the number of stress concentrators (dislocations and grain boundary effects) and contribution from the surface energy. The surface energy strongly increases during Si-I→Si-II PT, suppressing this PT for all particles. This effect increases with a reduction in particle size, and for 30 nm Si, Si-XI and V appear before Si-II due to their smaller surface energy. The evolution of phase fractions of Si particles under hydrostatic compression is presented for the first time. For 30 nm particles, Si-I coexists with Si-V up to 23.2 GPa. This is unexpected because the first principle simulations[36] produce transverse acoustic phonon instability at 18.3 GPa for perfect Si-I crystal, i.e., it cannot exist above this pressure. Consequently, Si-I is stabilized by size and surface effects. During full pressure release, the previously unreported reverse Si-V→Si-I PT is observed. For 30 nm particles, complete unloading leads to Si-V→ amorphous Si PT. Obtained results find applications for the determination of pressure distribution in non-hydrostatic compression and shear of Si[20] and for developing quantitative models for PT evolutions as outlined in[30].

**Acknowledgments:** We thankfully acknowledge the support from NSF (CMMI-1943710 and DMR-2246991) and Iowa State University (Vance Coffman Faculty Chair Professorship and Murray Harpole Chair in Engineering). The high-pressure experiments were performed at HPCAT (Sector 16), Advanced Photon Source (APS), and Argonne National Laboratory. HPCAT operations are supported by DOE-NNSA's Office of Experimental Science. The Advanced Photon Source is a U.S. Department of Energy (DOE) Office of Science User Facility operated for the DOE Office of Science by Argonne National Laboratory under Contract No. DE-AC02-06CH11357. We also acknowledge CDAC-UIC for helping with the laser drilling of gaskets.

**Author contributions:** SY and FL performed experiments. SY collected and analyzed the data. VIL conceived the study, supervised the project, and analyzed the results. KKP and JS assisted with experiments. SY and VIL prepared the manuscript.

**Competing interests**

The authors declare no competing interests.

**Data availability**

The data supporting this study's findings are available from the corresponding authors upon request.

## Appendix

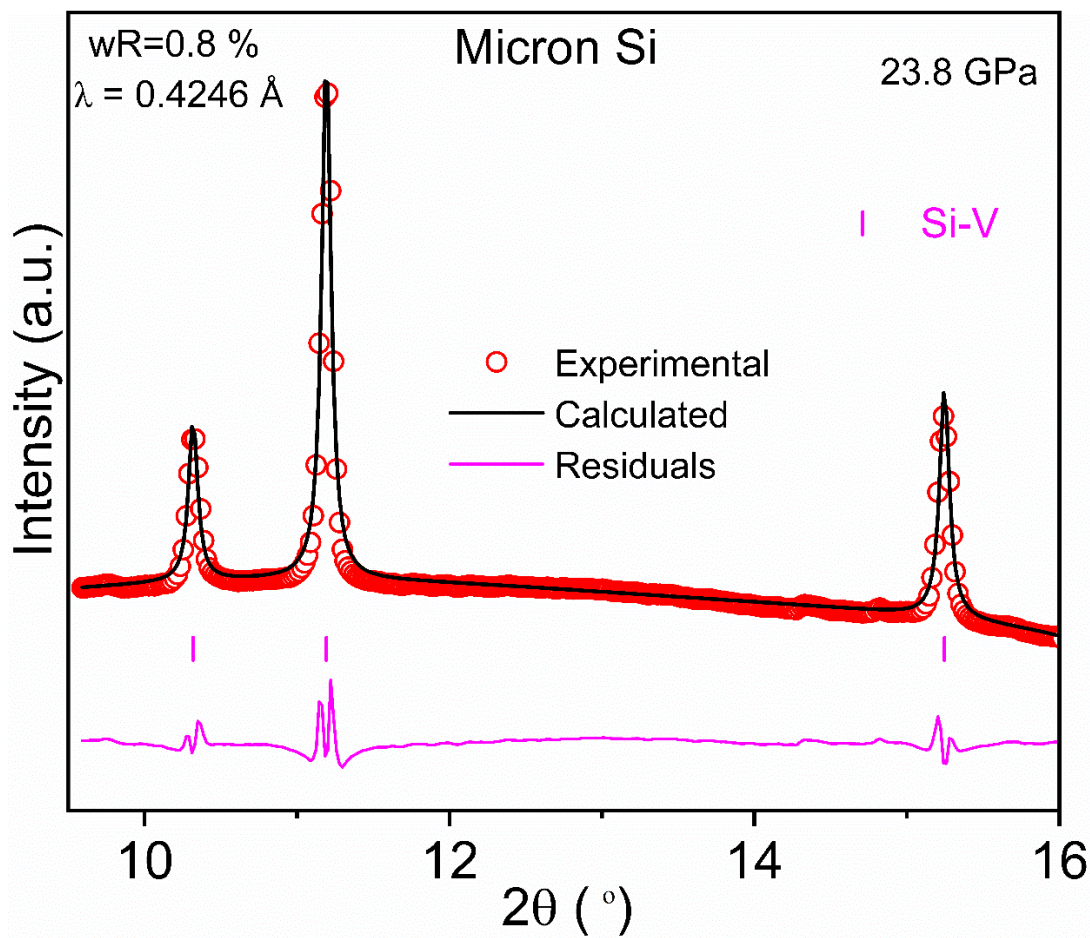

**Fig. A1.** Rietveld refinement of Si-V phase of micron Si at 23.8 GPa.

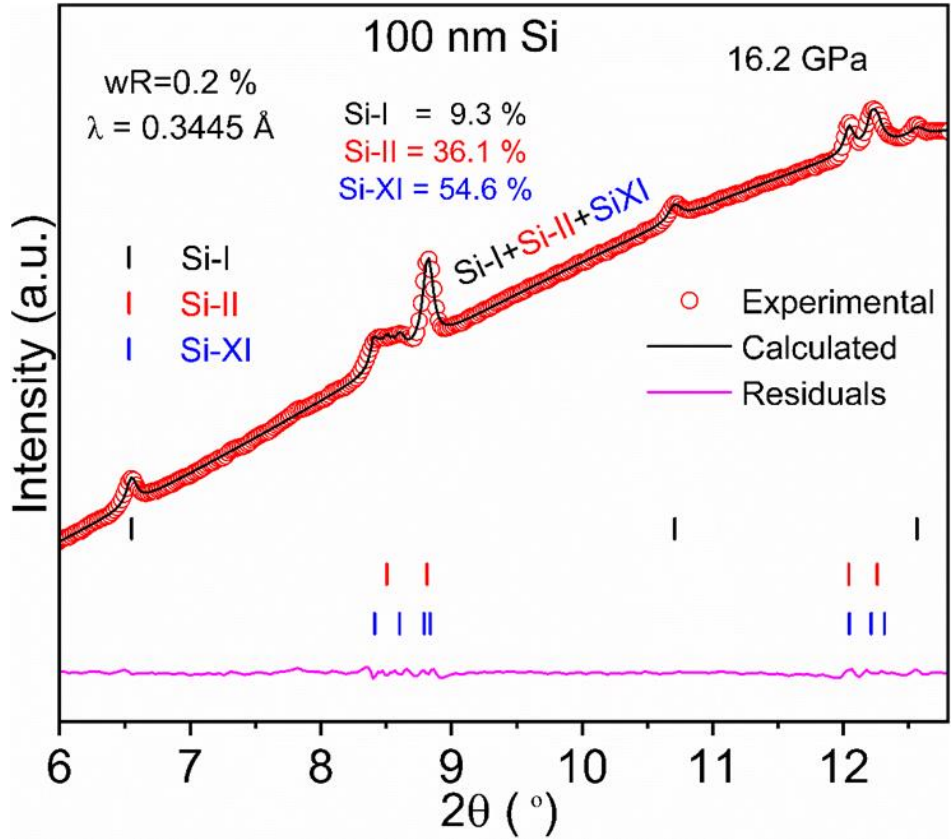

**Fig. A2.** Rietveld refinement of XRD patterns of 100 nm Si at 16.2 GPa. The Si-I, Si-II and Si-XI phases are coexisting. The phase fractions of all the phases are provided.